\documentclass{appolb}
\usepackage{epsfig}
\begin{document}
%\date{\today}
\pagestyle{plain}
%% uncomment the following line to get equations numbered by (sec.num)
%\eqsec
\newcount\eLiNe\eLiNe=\inputlineno\advance\eLiNe by -1
\title{ 
DIFFERENTIAL EQUATIONS for the 2-LOOP EQUAL MASS SUNRISE 
\thanks{Presented at the Ustron Conference, 15-21 September 2003; 
to be published in {\tt Acta Physica Polonica B.}} 
% \LaTeXe{} DOCUMENT CLASS\\
% FOR\\
% ACTA PHYSICA POLONICA B%
% \thanks{Send any remarks to {\tt acta@jetta.if.uj.edu.pl}}%
}
\author{ E. Remiddi 
         \address{ 
         Dipartimento di Fisica, Universit\`a di Bologna and 
         INFN, Sezione di Bologna, \\ 
         via Irnerio 46, I-40126 Bologna, Italy \\ 
         Institut f\"ur Theoretische Teilchen Physik, 
         Universit\"at Karslruhe, \\ 
         D-76128 Karslruhe, Germany 
         } 
} 
\maketitle  
\begin{abstract}
The differential equations for the 2-loop sunrise graph, at equal masses 
but arbitrary momentum transfer, are used for the analytic evaluation 
of the coefficients of its Laurent-expansion in the continuous dimension $d$. 
\end{abstract}
PACS: 12.15.Lk 

\vspace*{-10.5cm}{\hfill TTP-0326}

\vspace*{10cm} 

\section{Introduction}
\begin{figure}[h]
\begin{center}
\includegraphics*[2cm,0cm][10cm,4cm]{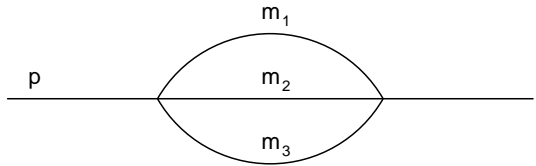}
\caption{\label{fig1} The 2-loop sunrise graph.}
\end{center}
% \label{fig:1}
\end{figure}

The differential equations in the squared external momentum $p^2$ 
for the Master Integrals (MIs) of the 2-loop sunrise graph with 
arbitrary masses $m_1, m_2, m_3$ of Fig.~\ref{fig1} were written 
in~\cite{CCLR1}. 
%%%%%%%%%%%%%%%%%%%%%%%%%%%%%%%%%%%%%%%%%%%%%%%%%%%%%%%%%%%%%%%%%%
%%%%%%%%%%%%%%%%%%%%%%%%%%%%%%%%%%%%%%%%%%%%%%%%%%%%%%%%%%%%%%%%%%
They were used for obtaining analytically particular values and 
behaviours at zero and infinite momentum transfer~\cite{CCLR1}, at 
pseudothresholds~\cite{CCR1} and threshold~\cite{CCR2}, as well 
as for direct numerical integration\cite{CCR3}. 
In this contribution I will report on some progress in the anlytic study 
of the solutions of the equations for arbitrary momentum transfer 
in the equal mass limit $m_i=1$. A more complete account 
will be given elsewhere~\cite{SLER}; while the algebraic burden 
in the arbitrary mass case will surely be much heavier, there 
are indications~\cite{HCER} that the approach can be extended to the 
arbitrary mass case as well. \par 
In the equal mass limit the 2-loop sunrise has two MIs, which in the 
usual $d$-continuous regularization scheme can be written as 
\begin{eqnarray} 
 S(d,p^2) &=& \frac{1}{\Gamma^2\left(3-\frac{d}{2}\right)} 
  \int\frac{d^dk_1}{ 4\pi^{ \frac{d}{2} } } 
  \int\frac{d^dk_2}{4\pi^{\frac{d}{2}}} 
  \frac{1}{(k_1^2+1)(k_2^2+1)[(p-k_1-k_2)^2+1]} \ , \nonumber\\ 
  \label{MIdef} \\ 
 S_1(d,p^2) &=& \frac{1}{\Gamma^2\left(3-\frac{d}{2}\right)} 
  \int\frac{d^dk_1}{4\pi^{\frac{d}{2}}} 
  \int\frac{d^dk_2}{4\pi^{\frac{d}{2}}} 
  \frac{1}{(k_1^2+1)^2(k_2^2+1)[(p-k_1-k_2)^2+1]} \ . \nonumber 
\end{eqnarray} 
Let us put $p^2=z$ ($z$ is positive when $p$ is Euclidean); the two 
MIs then satisfy the following linear system of first order 
differential equations in $z$ 
\begin{eqnarray} 
 z \frac{d}{dz} S(d,z) &=& (d-3) S(d,z) + 3 S_1(d,z) \ , \nonumber\\ 
 z(z+1)(z+9) \frac{d}{dz} S_1(d,z) &=& \frac{1}{2}(d-3)(8-3d)(z+3)S(d,z) 
 \label{1stosys} \\ 
  &+& \frac{1}{2} \left[ (d-4)z^2 + 10(2-d)z +9(8-3d) \right] S_1(d,z) 
                                                         \nonumber\\ 
  &+& \frac{1}{2} \frac{z}{(d-4)^2} \ . \nonumber 
\end{eqnarray} 
The system can be rewritten as a second order differential equation for 
$S(d,z)$ only 
\begin{eqnarray}
 z(z+1)(z+9) & \frac{d^2}{dz^2}S(d,z) & \nonumber\\
  + \frac{1}{2}\left[ (12-3d)z^2 + 10(6-d)z + 9d \right] & \frac{d}{dz}S(d,z)
                                                     &  \label{2ndoeq} \\ 
  + \frac{1}{2}(d-3) \left[ (d-4)z - d - 4 \right] & S(d,z) & =
    \frac{3}{2} \frac{1}{(d-4)^2} \ . \nonumber 
\end{eqnarray} 
As the second MI $S_1(d,z)$ can be written in terms of $S(d,z)$ and its 
first derivative 
\begin{equation} 
  S_1(d,z) = \frac{1}{3}\left[ - (d-3) + z\frac{d}{dz} \right] S(d,z) \ , 
\label{S1} 
\end{equation} 
we can take from now on $S(d,z)$ and its derivative $dS(d,z)/dz$ as the 
effective MIs. 

\section{From near 4 to near 2 dimensions. } 
We want to expand $S(d,z)$ around $d=4$ as Laurent series in $(d-4)$ 
and then to obtain analytically the values of the coefficients of 
the expansion by solving the 
relevant differential equations. It was found {\it a posteriori} that 
all the formuale are much simpler when expanding around $d=2$. 
To give the relations between the two expansions, let us recall that 
acting on any scalar Feynman integral in $d$ dimensions with a suitable 
differential operator, one obtains the same integral in $(d-2)$ 
dimensions. times a numerical factor depending on $d$ \cite{Tarasov}. 
Acting on the MIs in $d$ dimensions (or, in our case, on the two 
functions $S(d,z)$ and $dS(d,z)/dz$), one obtains the same MIs in $d-2$ 
dimensions in terms of mass derivatives of the MIs in $d$ dimensions, 
which can be expressed again in terms of MIs in $d$ dimensions; solving 
the linear system for the $d$-dimensional MIs and replacing finally 
$d$ by $d+2$ one obtains 
\begin{eqnarray}
 S(2+d,z) &=& \frac{1}{3(d-1)(3d-2)(3d-4)} \times \nonumber\\ 
   && \left\{ - \frac{9}{(d-2)^2} + \frac{3z-63}{4(d-2)} \right. 
      \label{d+2} \\ 
   && {\kern10pt} 
      + (z+1)(z+9) \left[ 1 + (z-3)\frac{d}{dz} \right] S(d,z) \nonumber\\ 
   && \left. \phantom{\frac{1}{1}}   % to force a big \right\} 
      + (d-2) (87+22z-z^2) S(d,z) \right\}  \nonumber 
\end{eqnarray} 

Quite in general, if 
\[ A(2+d) = B(d) \ , \] 
one can set $d=2+\eta$ and Laurent-expand in $\eta$; one obtains 
\[ \sum_k \eta^k A^{(k)}(4) = \sum_k \eta^k B^{(k)}(2) \ . \] 
\par 
The Laurent expansion in $\eta$ of $S(d,z)$ for $d=4+\eta$ begins with 
a double pole in $\eta$ and reads 
\[ S(4+\eta,z) = \frac{1}{\eta^2} S^{(-2)}(4,z) 
               +  \frac{1}{\eta} S^{(-1)}(4,z) 
               + S^{(0)}(4,z) + \eta S^{(1)}(4,z) + ....... \] 
while $S(2+\eta,z)$ has no singularities in $\eta$, and its expansion is 
\[ S(2+\eta,z) = S^{(0)}(2,z) + \eta S^{(1)}(2,z) + ....... \] 
By inserting the two expansions in 
Eq.(\ref{d+2}), one gets the required coefficients $S^{(k)}(4,z)$ of the 
Laurent expansion in $\eta$ of $S(4+\eta,z)$ around $4$ in terms of 
the coefficients $S^{(k)}(2,z)$ of the expansion of $S(2+\eta,z)$ 
around $2$. 
As $S(d,z)$ is regular at $d=2$, the poles in $\eta$ 
are not hidden in $S(d,z)$ but are explicitly exhibited by the 
$1/(d-2)$ factors in the r.h.s. of Eq.(\ref{d+2}). 
Working out the algebra, one finds at once 
\begin{eqnarray} 
    S^{(-2)}(4,z) &=& - \frac{3}{8} \ , \nonumber\\ 
    \label{polesat4} \\ 
    S^{(-1)}(4,z) &=& \frac{9}{16} + \frac{z}{32} \ . \nonumber 
\end{eqnarray}

\section{The expansion at $d=2$ of the differential equation.} 
By expanding systematically in $(d-2)$ all the terms appearing in 
Eq.(\ref{2ndoeq}), one obtains a set of chained equations of the 
form 
\begin{eqnarray}
 \biggl\{ \ \ \frac{d^2}{dz^2} +
         \left[\frac{1}{z} + \frac{1}{z+1} + \frac{1}{z+9}
        \right] \frac{d}{dz} & & \nonumber\\
      \label{chain} \\ 
      + \left[ \frac{1}{3z} - \frac{1}{4(z+1)} - \frac{1}{12(z+9)}
         \right] & \biggr\} S^{(k)}(2,z) = N^{(k)}(2,z) \nonumber 
\end{eqnarray} 
where the homogeneous part is the same for any order $k$, and the first 
few inhomogeneous terms are 
\begin{eqnarray} 
 N^{(0)}(2,z) &=& \frac{1}{24z} - \frac{3}{64(z+1)} + \frac{1}{192(z+9)} 
               = \frac{3}{8z(z+1)(z+9)} \ ,           \nonumber\\ 
 N^{(1)}(2,z) &=& \left( - \frac{1}{2z} + \frac{1}{z+1} + \frac{1}{z+9} 
                  \right)\frac{dS^{(0)}(2,z)}{dz} \label{Nchained} \\ 
              &+& \left( \frac{5}{18z} - \frac{1}{8(z+1)} 
                   - \frac{11}{72(z+9)} \right) S^{(0)}(2,z) \nonumber\\ 
              &+& \frac{1}{24z} - \frac{3}{64(z+1)} + \frac{1}{192(z+9)} 
                                                  \ , \nonumber\\ 
 N^{(2)}(2,z) &=& ..... \ . \nonumber 
\end{eqnarray} 
The equations Eq.(\ref{Nchained}) are chained, in the sense that the 
inhomogeneous term of order $k$ involves lower terms, of order $(k-1)$ 
(for $k>0$) and $(k-2)$ (for $k>1$) in the expansion of $S(2,z)$, 
as can be seen from Eq.(\ref{2ndoeq}) and is shown explicitly in 
Eq.s(\ref{Nchained}). 
\par 
The system Eq.(\ref{chain}) is 
to be solved bottom up in $k$, starting from $k=0$ (in which 
case the inhomogeneous term is completely known) and then proceeding 
to higher values increasing $k$ by one, so that at each step the 
inhomogeneous term is known from the solution of the previous equations. 
The chained equations can then be solved by using Euler's method of the 
variation of the constants. The homogeneous equation is the same for 
all the values of $k$, 
\begin{eqnarray} 
 \biggl\{ \ \ \frac{d^2}{dz^2} + 
         \left[\frac{1}{z} + \frac{1}{z+1} + \frac{1}{z+9} 
        \right] \frac{d}{dz} & & \nonumber\\ 
        \label{homo} \\ 
      + \left[ \frac{1}{3z} - \frac{1}{4(z+1)} - \frac{1}{12(z+9)} 
         \right] & \biggr\} \Psi(z) = 0 \ ; \nonumber 
\end{eqnarray} 
if $\Psi_1(z), \Psi_2(z)$ are two independent solutions of the homogeneous 
equation, $W(z)$ the corresponding Wronskian 
\begin{equation} 
  W(z) = \Psi_1(z) \frac{d\Psi_2(z)}{dz} 
        - \Psi_2(z) \frac{d\Psi_1(z)}{dz} 
\label{Wronskian} 
\end{equation} 
according to Euler's method the solutions of Eq.s(\ref{chain}) are 
given by the integral representations 
\begin{eqnarray} 
 S^{(k)}(2,z) &=& \Psi_1(z) \left( \Psi^{(k)}_1 
     - \int_0^z \frac{dw}{W(w)} \Psi_2(w) N^{(k)}(2,w) \right) \nonumber\\ 
       \label{Euler} \\ 
              &+& \Psi_2(z) \left( \Psi^{(k)}_2 
     + \int_0^z \frac{dw}{W(w)} \Psi_1(w) N^{(k)}(2,w) \right) \ , \nonumber 
\end{eqnarray} 
where \( \Psi^{(k)}_1, \Psi^{(k)}_2 \) are two integration constants. 
\par 
Eq.(\ref{Euler}) at this moment is just a formal representation of the 
solutions for the coeffcients $S^{(k)}(2,z)$; it becomes a 
substancial (not just formal!) formula only when all the ingredients 
-- the two solutions of the homogeneous equation $\Psi_i(z)$, their 
Wronskian $ W(z) $ and the two integration constants $ \Psi^{(k)}_i $ 
are known explicitly. 
\par 
Although the Wronskian is defined in terms of the $\Psi_i(z)$, it can be 
immediately obtained (up to a multiplicative constant) from 
Eq.(\ref{homo}). An elementary calculation using the definition 
Eq.(\ref{Wronskian}) and the value of the second derivatives of the 
$\Psi_i(z)$, as given by Eq.(\ref{homo}) of which they are solutions, 
leads to the equation 
\[ \frac{d}{dz} W(z) = - \left( \frac{1}{z} + \frac{1}{z+1}
                       + \frac{1}{z+9} \right) W(z) \ , \]
which gives at once 
\begin{equation} 
 W(z) = \frac{9}{z(z+1)(z+9)} \ , \label{Wvalue} 
\end{equation} 
where the multiplicative constant has been fixed anticipating later 
results. 
\par 
Finding the two $\Psi_i(z)$ requires much more work. 

\section{Solving the homogeneous equation at the singular points.} 
By inspection, the singular points of Eq.(\ref{homo}) are found to be 
\[  z=0, -1, -9, \infty \ ; \]
at each of those points one has two independent solutions, the first 
regular and the second with a logarithmic singularity. 
The expansions of the solutions around each of the singular points is 
immediately provided by the differential equation itself. 
\par 
Around $z=0$ the two solutions of Eq.(\ref{homo}) can be written as 
\begin{eqnarray} 
  \Psi_1^{(0)}(z) &=& \psi_1^{(0)}(z) \nonumber\\ 
  \Psi_2^{(0)}(z) &=& \ln{z} \ \psi_1^{(0)}(z) + \psi_2^{(0)}(z) \ , 
\label{Psiat0} 
\end{eqnarray} 
where the $\psi_i^{(0)}(z)$ are power series in $z$. 
Imposing $\psi_1^{(0)}(0)=1$, one finds 
\begin{eqnarray} 
 \psi_1^{(0)}(z) &=& 1 - \frac{1}{3}z + \frac{5}{27}z^2 + ... \ ,\nonumber\\ 
 \psi_2^{(0)}(z) &=& - \frac{4}{9}z + \frac{26}{81}z^2 + ...  \ ; 
\label{Psiat0a} 
\end{eqnarray} 
the coefficients are given recursively (hence up to any reuired order) 
by the equation. The radius of convergence is 1 (the next singularity is 
at $z=-1$) and the two solutions are real for positive $z$ (spacelike 
momentum transfer). The continuation to the timelike region 
is done by giving to $z$ the value $z=-(u+i\epsilon)$; 
for $ 0< u < 1\, $ one has $\ln{z} = \ln{u} - i\pi $ and 
$ \Psi_2^{(0)}(z) $ develops an imaginary part $-i\pi \psi_1^{(0)}(z)$. 
\par 
Similarly, around $z=-1$ the 2 independent solutions can be written as 
\begin{eqnarray} 
  \Psi_1^{(1)}(z) &=& \psi_1^{(1)}(z) \nonumber\\ 
  \Psi_2^{(1)}(z) &=& \ln{(z+1)} \psi_1^{(1)}(z) + \psi_2^{(1)}(z) \ , 
\label{Psiat1} 
\end{eqnarray} 
with Eq.(\ref{homo}) providing recursively the coefficients of the 
expansions in powers of $(z+1)$ of the two $\psi_i^{(1)}(z)$ once the 
initial condition is given. If $\psi_1^{(1)}(0)=1 $ one has 
\begin{eqnarray} 
 \psi_1^{(1)}(z) &=& 1 + \frac{1}{4}(z+1) + \frac{5}{32}(z+1)^2 + ... \ , 
                                                             \nonumber\\ 
 \psi_2^{(1)}(z) &=& + \frac{3}{8}(z+1) + \frac{33}{128}(z+1)^2 + ... \ , 
\label{Psiat1a} 
\end{eqnarray} 
with radius of convergence 1 (up to the singularity at $z=0$) etc. 
\par 
The other two singular points $z=-9$ and $z=\infty$ can be treated in the 
same way, the corresponding formualae are not given due to lack of space. 
\par 

\section{The interpolating solutions.} 
Having the solutions piecewise is not sufficient, one must build two 
solutions in the whole $ -\infty < z < \infty $ range by suitably 
joining the above expressions of the solutions at the singular points. 
A hint is provided by the knowledge of the imaginary part of the 
original Feynman integral $S(d,p^2)$ Eq.(\ref{MIdef}) at $d=2$ 
dimensions; as already observed in~\cite{GP}, 
the Cutkosky-Veltman rule gives for the imaginary part of $S(d,p^2)$ 
at $d=2$ and $ u = -z \ge 9 $ (and up to a multiplicative constant) the 
integral representation
\begin{equation} 
  J(u) = \int_4^{(\sqrt{u}-1)^2} \frac{db}{\sqrt{R_4(u,b)} } \ , 
\label{ImS} 
\end{equation} 
where $ R_4(u,b) $ stands for the polynomial (of 4th 
order in $b$ and 2nd order in $u$) 
\[ R_4(u,b) = b(b-4)(b-(\sqrt{u}-1)^2) (b-(\sqrt{u}+1)^2) \ , \] 
and the $b$ integration runs between two adjacent zeros of $ R_4(u,b) $. 
As the inhomogeneous part of Eq.(\ref{2ndoeq}) cannot develop an imaginary 
part, the imaginary part of the Feynman integral in $d=2$ dimensions, 
$J(u)$ of Eq.(\ref{ImS}), is necessarily a solution of the associated 
homogeneous equation at $d=2$, \ie of Eq.(\ref{homo}) -- a fact which 
can also be checked explicitly. 
\par 
One is then naturally lead to consider all the $b$-integrals of 
$1/\sqrt{R_4(u,b)}$ between any two adjacent roots for all possible values 
of $u$. 
The details of the analysis cannot be reported here again for lack of 
space. As a result, one finds for instance that when $u$ is in the range 
$0<u<1$ the roots of $ R_4(u,b) $ are ordered as 
\[ 0<(\sqrt{u}-1)^2<(\sqrt{u}+1)^2<4 \] 
and the associated $b$-integrals are 
\begin{eqnarray} 
  J_1^{(0,1)}(u) &=& \int_0^{(\sqrt{u}-1)^2} \frac{db}{\sqrt{-R_4(u,b)}} 
                                             \ , \nonumber\\ 
  J_2^{(0,1)}(u) &=& \int_{(\sqrt{u}-1)^2}^{(\sqrt{u}+1)^2} 
                 \frac{db}{\sqrt{R_4(u,b)}}  \ , \nonumber\\ 
  J_3^{(0,1)}(u) &=& \int_{(\sqrt{u}+1)^2}^4 
                 \frac{db}{\sqrt{-R_4(u,b)}} \ . \nonumber 
\end{eqnarray} 
The three integrals are all real (and positive) due to the choice of the 
sign in front of $R_4(u,b)$ in the square roots; more important, 
they all satisfy Eq.(\ref{homo}) -- therefore they cannot be all 
independent. With standard changes of variables, they can be brought 
in the form of Legendre's complete elliptic integrals of the first 
kind~\cite{GP}; in that way one finds for instance 
\[ J_1^{(0,1)}(u) = J_3^{(0,1)}(u) 
                  = \frac{1}{\sqrt{(1+\sqrt{u})^3(3-\sqrt{u})}} 
     K\left(\frac{(1-\sqrt{u})^3(3+\sqrt{u})} 
                 {(1+\sqrt{u})^3(3-\sqrt{u})}\right) \ , \] 
showing in particular that the first and third integrals are indeed 
equal. A similar (but different!) formula holds for the second integral. 
\par 
Although usually one can not do very much for expressing elliptic integrals 
in terms of other more familiar functions (such as logarithms or the like), 
in the limiting cases $u\to 0$ and $u \to 1$ two of the 4 roots of 
$R_4(b,u)$ become equal, and the by now elementary $b$-integrations give 
\begin{eqnarray} 
 \lim_{u\to 0^+} J_1^{(0,1)}(u) &=& \frac{1}{\sqrt3} 
          \left( - \frac{1}{2}\ln{u} + \ln3 \right) \ ,   \nonumber\\ 
 \lim_{u\to 0^+} J_2^{(0,1)}(u) &=& \frac{\pi}{\sqrt3} \ , 
          \label{J01at0} \\ 
 \lim_{u\to 1^-} J_1^{(0,1)}(u) &=& \frac{\pi}{4}   \ ,   \nonumber\\ 
 \lim_{u\to 1^-} J_2^{(0,1)}(u) &=& -\frac{3}{4}\ln(1-u) 
                                 + \frac{9}{4}\ln2  \ . 
          \label{J01at1} 
\end{eqnarray} 
One has now all the information needed for defining two solutions 
$\Psi_i(z)$ of Eq.(\ref{homo}) in an interval which contains the 
two singular points $z=0$ and $z=-1$. 
Let us start by defining, for $z>0$, 
\begin{eqnarray} 
  \Psi_1(z) &=& \Psi_1^{(0)}(z) \ , \nonumber\\ 
  \Psi_2(z) &=& \Psi_2^{(0)}(z) \ . \nonumber
\end{eqnarray} 
That fixes the multiplicative constant in the Wronskian as well, 
giving the result already anticipated in Eq.(\ref{Wvalue}). 
From Eq.s(\ref{Psiat0},\ref{Psiat0a}) we easily read the behaviours 
of the $\Psi_i(z)$ for 
$u=-z$ small and positive; but in the range $0<u<1$ the solutions can also 
be expressed in terms of the $J_i^{(0,1)}(u)$; by matching the behaviours 
$u\to 0^+$ of the $J_i^{(0,1)}(u)$, Eq.s(\ref{J01at0}) to the 
behaviours of the $ \Psi_i(z) $, one finds, in the interval $0<u<1$, 
\begin{eqnarray} 
 \Psi_1(z-i\epsilon) &=& \frac{\sqrt3}{\pi} J_2^{(0,1)}(u) \ , \nonumber\\ 
 \Psi_2(z-i\epsilon) &=& -2\sqrt3  J_1^{(0,1)}(u) 
          + \frac{\sqrt3}{\pi} (2\ln3-i\pi) J_2^{(0,1)}(u) \ . 
\label{PsiasJ01} 
\end{eqnarray} 
One can now compare Eq.s(\ref{J01at1}) with Eq.s(\ref{Psiat1},\ref{Psiat1a}) 
and express the $J_i^{(0,1)}(u)$ in terms of the $ \Psi_i^{(1)}(z) $; 
substituting in Eq.(\ref{PsiasJ01}) one finds for the solutions 
$ \Psi_i(z) $, for $z$ around $-1$, the values 
\begin{eqnarray}
 \Psi_1(z-i\epsilon) &=& \frac{9\sqrt3}{4\pi}\ln2\ \Psi_1^{(1)}(z-i\epsilon)
                      - \frac{3\sqrt3}{4\pi} \Psi_2^{(1)}(z-i\epsilon) \ , 
                                                       \nonumber\\ 
 \Psi_2(z-i\epsilon) &=& \frac{\sqrt3}{4}\left( \frac{18}{\pi}\ln2\ln3 
          - 2\pi - i9\ln2 \right) \Psi_1^{(1)}(z-i\epsilon) \nonumber\\ 
       &+& \frac{3\sqrt3}{4\pi}(-2\ln3+i\pi) \Psi_2^{(1)}(z-i\epsilon) \ . 
          \nonumber 
\end{eqnarray} 
One can then move to the next interval $ 1 < u < 9 $ and so on till the 
$ \Psi_i(z) $ are expressed, in the whole range $ -\infty < z < \infty $, 
in terms of the $ \Psi_i^{(k)}(z) $, each known 
within the convergence radius of the expansions given by Eq.(\ref{homo}), 
as well as in terms of the elliptic integrals $J_i^{(k,l)}(u)$. 

\section{The integration constants.} 
$S(d,p^2)$ Eq.(\ref{MIdef}) is known to be real for $u=-p^2$ below the 
threshold at $u=9$. Take the solution as given by Euler's formula 
Eq.(\ref{Euler}); in the region $0<u<1$, or $0>z>-1$ the argument 
$w$ of the inhomogeneous term $N^{(k)}(w)$ varies in the interval 
$0>w>z>-1$, and is therefore real ($N^{(k)}(w)$, Eq.(\ref{Nchained}) 
involves either real algebraic fractions or lower order terms of the 
expansion in $(d-2)$ of $S(d,z)$, which are real in that region) -- 
therefore an imaginary part, if any, can come only from the $\Psi_i(z)$ 
and the $\Psi_i(w)$. By using the values of the $\Psi_i(z)$ as given 
by Eq.s(\ref{PsiasJ01}), one finds for $u=-z$ in the range $0<u<1$ 
\[ \mbox{\rm Im}S^{(k)}(2,z) = - \sqrt{3}\Psi_2^{(k)} J_2^{(0,1)}(u) \]
implying , for any $k$, 
\[ \Psi_2^{(k)} = 0 \ . \] 
\par 
The argument can be repeated in the interval $1<u<9$ (between 
pseudothreshold and threshold), where the $\Psi_i(z)$, 
in analogy with Eq.s(\ref{PsiasJ01}), are expressed in terms of 
\begin{eqnarray} 
  J_1^{(1,9)}(u) &=& \int_0^{(\sqrt{u}-1)^2} \frac{db}{\sqrt{-R_4(u,b)}} 
                                             \ , \nonumber\\ 
  J_2^{(1,9)}(u) &=& \int_{(\sqrt{u}-1)^2}^4 
                 \frac{db}{\sqrt{R_4(u,b)}}  \ . \nonumber 
\end{eqnarray} 
One finds for $z$ in the interval $ -1 > z > -9,\ $ \ie $\ 1<u<9$ 
\[ \mbox{\rm Im}S^{(k)}(2,z) = 3\frac{\sqrt3}{\pi} 
     J_1^{(1,9)}(-z)\left( \Psi_1^{(k)} 
   + 2\sqrt3\int_0^{-1}\frac{dw}{W(w)} J_1^{(0,1)}(-w) N^{(k)}(2,w) 
                                                 \right) \ ; \] 
as the imaginary part must vanish in that interval, the other integration 
constant is given by 
\[ \Psi_1^{(k)} =
   - 2\sqrt3\int_0^{-1}\frac{dw}{W(w)} J_1^{(0,1)}(-w) N^{(k)}(2,w) \ . 
\]

\section{Conclusions.} 
The values of all the quantities entering in Eq.(\ref{Euler}), 
namely the two $\Psi_i(z),\ W(z)$ and the two integration constants 
$K_i^{(k)}$ have been obtained, so that the previously formal 
expression given by Eq.(\ref{Euler}) became a substancial formula 
giving the functions $S^{(k)}(2,z)$ in closed analytic form. 
Indeed, from the explicit knowledge of the singularities and the relevant 
expansions of the $\Psi_i(z)$, the singularities and the relevant 
expansions of the $S^{(k)}(2,z)$ are immediately obtained -- and from the 
explicit knowledge of the singularities and the relevant expansions 
of the $S^{(k)}(2,z)$ the fast and precise numerical 
routines for their evaluation can in turn be obtained. 

\vskip 1cm \noindent 
{\bf Acknowledgements.} The author wants to thank J.~Vermaseren for his 
continuous and kind assistance in the use of his algebra manipulating 
program {\tt FORM}~\cite{FORM}, by which all the calculations were 
carried out.

%%%%%%%%%%%%%%%%%%%%%%%%%%%%%%%%%%%%%%%%%%%%%%%%%%%%%%%%%%%%%%%%%%%%%%%% 

\def\NP{{\sl Nucl. Phys.}} 
\def\PL{{\sl Phys. Lett.}} 
\def\PR{{\sl Phys. Rev.}} 
\def\PRL{{\sl Phys. Rev. Lett.}} 
\def\NC{{\sl Nuovo Cim.}}
\def\APP{{\sl Acta Phys. Pol.}}
\def\ZP{{\sl Z. Phys.}}
\def\MPL{{\sl Mod. Phys. Lett.}} 
\def\EPJ{{\sl Eur. Phys. J.}} 
\def\IJMP{{\sl Int. J. Mod. Phys.}}

\end{document}